\documentclass[aps,prl,reprint,groupedaddress]{revtex4-1}
\usepackage[dvipdfm,colorlinks,pdfstartview=FitH,citecolor=blue,linkcolor=blue,urlcolor=blue]{hyperref}
\usepackage{amsmath}
\usepackage{bm}
\usepackage{braket}
\usepackage{graphicx}

\begin{document}


\title{Analytical Perspective for Bound States in the Continuum in Photonic Crystal Slabs}


\author{Yi Yang$^1$}
\author{Chao Peng$^1$}
\email{pengchao@pku.edu.cn}
\author{Yong Liang$^2$}
\author{Zhengbin Li$^1$}
\author{Susumu Noda$^2$}
\affiliation{$^1$State Key Laboratory of Advanced Optical Communication Systems $\&$ Networks, Department of Electronics, Peking University, Beijing 100871, China\\
$^2$Department of Electronic Science and Engineering, Kyoto University, Kyoto-Daigaku-Katsura, Nishikyo-ku, Kyoto 615-8510, Japan}

\date{\today}

\begin{abstract}
We investigate the formation of photonic bound states in the continuum (BICs) in photonic crystal slabs from an analytical perspective. Unlike the stationary at-$\Gamma$ BICs which origin from the geometric symmetry, the tunable off-$\Gamma$ BICs are due to the weighted destructive via-the-continuum interference in the vicinity of accidental symmetry when the majority of the radiation is pre-canceled. The symmetric compatible nature of the off-$\Gamma$ BICs leads to a trapping of light that can be tuned through continuously varying the wavevector. With the analytical approach, we explain a reported experiment and predict the existence of a new BIC at an unrevealed symmetry.
\end{abstract}


\maketitle

The localization of waves has always been more difficult to manipulate than their propagation. It is well known that an electromagnetic wave of a specific frequency can be trapped by structures such as photonic \cite{DEFNAT,CAVNAT1,NewNP} and plasmonic \cite{CAVNAT3} nanocavities, in which outgoing waves are completely forbidden. However, it has been demonstrated that perfect light confinement can still be achieved even with allowed outgoing waves, because of a particular type of localized state: a bound state in the continuum (BIC). Historically, von Neumann and Wigner \cite{von} first proposed that a BIC can be explicitly constructed in a quantum system when the wave function exhibits weakly damped oscillations. Furthermore, the occurrence of BICs was interpreted as the interference of resonances in direct and via-the-continuum channels \cite{Feshbach,PRAFriedrich}. It is a very general effect that is important in many areas of physics, including photonics \cite{Pho1,Pho2,MolinaPRL,SadreevPho}, quantum \cite{Qu1,Asaf,Qu2,Qu3,SadreevQua}, aquatic and acoustic waves \cite{Wa1,Wa2}, \textit{etc.}. Although some artificially designed potentials cannot be readily realized in an electronic quantum system, similar phenomena may be more easily implemented optically under Maxwell's theory. Recently, BICs have been demonstrated in photonic crystal (PhC) slabs \cite{GMRPRL}, in which periodic geometry leads to BICs at $\Gamma$ on photonic band structures that are analogous to electron band structures in solids.

In many reported cases, BICs are decoupled from continuum states because geometric symmetry forbids coupling to any outgoing wave, i.e., symmetry incompatibility that leads to stationary BICs, namely, fixed BICs at $\Gamma$ point.
Very recently, another type of BIC was reported in PhC slabs on TM-like bands \cite{GMRNAT,LSA} at some seemingly unremarkable wavevectors without symmetry incompatibility, giving rise to a tunable trapping of light. This disappearance of leakage was preliminarily attributed to the destructive interference among several leakage channels, but the underlying physics remains unrevealed.

For a general quantum system, a theory regarding BICs was developed by Feshbach, Friedrich, and Wintgen \cite{Feshbach,PRAFriedrich} many years ago. It describes two interfering resonances caused by the coupling in closed and open channels. Similar phenomena also occur in a PhC slab system in which several guided mode resonances interfere. These resonances are coupled to each other through in-plane Bloch waves (closed channels) or leaky waves (open channels) and form the BICs. Recently, we proposed a coupled-wave theory (CWT) \cite{LCPRB,LCPRB2,HCCWT,TMCWT} for analyzing the guided mode resonance in two-dimensional PhC slabs. The CWT depicts the analogous physics in a photonic system in great detail, as what Feshbach's theory for a quantum system. In this Letter, we present a more comprehensive understanding of BICs in PhC slabs based on the analytical CWT of PhC modes in the continuum.
\begin{figure}[htp]
\centerline{
\includegraphics[width=8cm]{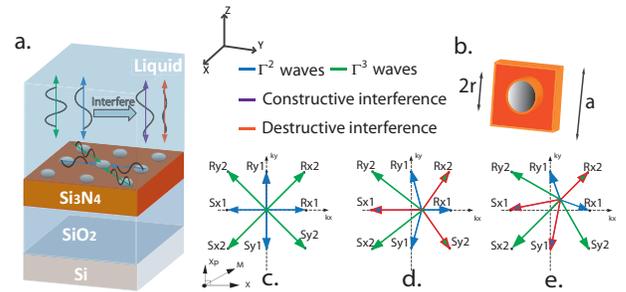}}
 \caption{(Color online). \textbf{a.} Structure: liquid ($\varepsilon_{l}=2.11$),  Si$_{3}$N$_{4}$ PhC slab ($\varepsilon_{sn}=4.08$), silica cladding ($\varepsilon_{so}=2.13$), and Si substrate. Setting $\varepsilon_{l}\simeq\varepsilon_{so}$ ensures mirror-flip symmetry \cite{GMRNAT}. The silica and liquid layers are sufficiently thick to be assumed as infinite.
 \textbf{b.} Basic lattice: thickness $2d = 180$ nm, periodicity $a =336$ nm, hole diameter $2r = 160$ nm. \textbf{c,d,e.} Phase matching conditions: \textbf{c.} $\Gamma$ point; \textbf{d.} $k=0.25$ in the $\Gamma-X$ direction; \textbf{e.} $k=\sqrt{2}/6$ in the $\Gamma-M$ direction. Red arrows indicate extra degenerate wavevectors. For square lattice, the $\Gamma^1$, $\Gamma^2$, and $\Gamma^{3}$ points are defined as wavevectors with lengths of 0, $2\pi/a$, and $2\sqrt{2}\pi/a$, respectively \cite{VecBeam}.
}
\label{schematic}
\end{figure}

A tunable BIC has been observed in the structure illustrated in Fig. \ref{schematic}a \cite{GMRNAT} and we focus on this structure for consistency. Assuming the TE-TM coupling is much weaker than their own internal coupling, the TM-like electromagnetic field is given by $\left(H_{x},H_{y},E_{z}\right)$. For a macroscopic PhC slab, assuming that the area in the \textbf{xy} plane is infinite, we have $H_i(z)=H_{i,0,0}+\sum_{m\neq 0,n\neq 0}H_{i,m,n}(z)e^{-im\beta_0 x-in\beta_0 y}$, where $\beta_{0}=2\pi/a$. Moreover, within the PhC slab ($|z|<d$), $1/\varepsilon(r)$ can be expanded with the Fourier transform: $1/\varepsilon(r) = \kappa_{a}+\sum\kappa_{mn}e^{-im\beta_0 x-in\beta_0 y}$, where $\kappa_a = f(1/\varepsilon_{l})+(1-f)(1/\varepsilon_{sn})$ ($f$ is the filling factor). Outside the PhC slab ($|z|>d$), $1/\varepsilon(r) = 1/\varepsilon_{l}\triangleq \kappa_{b}$. The coupling equations of closed and open channels can be obtained from Maxwell's equations, as follows for the \textbf{x} direction \cite{TMCWT}
\begin{widetext}
\begin{eqnarray}
\begin{split}
&\left(\kappa_{a}\frac{\partial^2}{\partial z^2}+\delta\left(\left|d\right|\right)\Delta\kappa\frac{\partial}{\partial z}+k_{0}^2-\kappa_{a}n_{y}^2\beta_{0}^{2}\right)H_{x,m,n}+\kappa_{a}m_{x}n_{y}\beta_{0}^{2}H_{y,m,n}  \\
&=\sum_{m'\neq m,n'\neq n}\kappa_{\substack{m-m',\\n-n'}}\left[\left(-\frac{\partial^2}{\partial z^2}
+\delta\left(\left|d\right|\right)\frac{\partial}{\partial z}+n_{y}n_{y}'\beta_{0}^{2}\right)H_{x,m',n'}-m_{x}'n_{y}\beta_{0}^{2}H_{y,m',n'}\right],
\label{CX}
\end{split}
\end{eqnarray}
\end{widetext}
where $\Delta \kappa \triangleq \kappa_{b}-\kappa_{a}$, $\delta\left(\left|d\right|\right) \triangleq \delta(z-d) -\delta(z+d)$ and $\left(m_x,n_y\right)$ follow the same notations as that used in our previous work \cite{LCPRB2}. Because of symmetry, the equation for the \textbf{y} direction can be readily obtained from Eq. \ref{CX} by switching $H_{x,m,n}$, $H_{x,m',n'}$, $m_{x}$, and $m_{x}'$ with $H_{y,m,n}$, $H_{y,m',n'}$, $n_{y}$, and $n_{y}'$, respectively. In the \textbf{z} direction, the transverse wave condition of the $\textbf{H}$ components yields
\begin{eqnarray}
\sum_{m',n'}\kappa_{\substack{m-m',\\n-n'}}\frac{\partial }{\partial z}\left(m_{x}H_{x,m',n'}+n_{y}H_{y,m',n'}\right)=0.
\label{CZ}
\end{eqnarray}

Because the TM-like modes involve a longitudinal electric field, two effects influence the coupling strength between individual channels: the in-plane coupling caused by the permittivity periodicity and the surface coupling caused by the discontinuities at dielectric interfaces. Essentially different from TE-like modes, the $\delta\left(\left|d\right|\right)$ part of the operator in Eq. \ref{CX} depicts the unique surface coupling existing in TM-like modes only \cite{yamamotoTM}. The remainder of the operator represents the conventional in-plane coupling for both TE and TM-like modes \cite{yamamotoTE,yamamotoTM}. Eq. \ref{CX} is analogous to Eq. 1 in Friedrich's work for a quantum system that describes the interference of several closed channels with one or more open channel(s) \cite{PRAFriedrich}.

The guided resonance depends on the phase matching between the guided mode ($\beta$) and a given $(m,n)$ order of the Bloch mode ($\beta_{mn}=m\beta_0 \hat x + n\beta_0 \hat y$), with $\beta=|\beta_{mn}|$. At $\Gamma^2$ point, with the phase matching given by $\beta=\beta_{0}$,
all Bloch waves except $H_{0,0}$ are confined within the slab.
In Fig \ref{schematic}c, owing to the symmetry of a square lattice, four wavevectors $\textbf{V}=\{R_{x1}, S_{x1}, R_{y1}, S_{y1}\}$ can be treated as uncoupled closed channels.
The coupling between closed channels can be realized via in-plane waves (through $H_{m,n}$), or via the continuum (through $H_{0,0}$). Thus we have an eigenvalue problem:
\begin{eqnarray}\label{CE}
\left(\textbf{k}-\textbf{k}_{0;mn}\right)\textbf{V}=&&\left(\sum_{m,n;m',n'}\bra{H_{i,m,n}}\right.
\kappa_{\substack{m-m',\\n-n'}}\ket{H_{i,m',n'}}\nonumber \\
&&\left.+\sum_{\substack{m,n}}\bra{H_{i,m,n}}\kappa_{\substack{m,n}}\ket{H_{i,0,0}}\right)\textbf{V}.
\end{eqnarray}
where $i=x, y$. Using the approach similar to that for TE-like modes \cite{LCPRB}, all terms in Eq.\ref{CE} can be solved analytically and the complex frequencies are obtained as eigenvalues.
\begin{figure}[htbp]
\centerline{
\includegraphics[width=8cm]{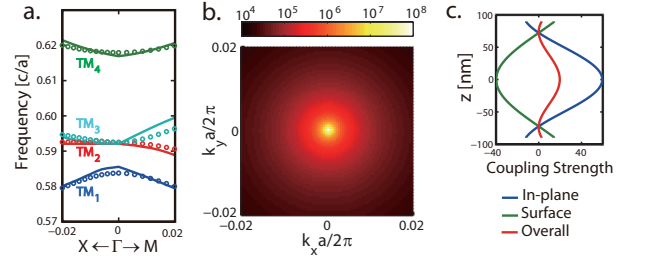}}
 \caption{(Color online). \textbf{a.} Identified TM$_{1-4}$ modes near $\Gamma^2$ point by increasing frequency  \cite{VecBeam}, using both CWT (lines) and FDTD (dots). \textbf{b.} CWT-calculated quality factor $Q_{r}$ of TM$_{1}$ mode near $\Gamma^2$. \textbf{c.} Pre-cancellation within a closed channel at $\Gamma^2$. The coupling strength \cite{LCPRB} is defined as $\bra{G}\kappa_{m,n}{L}\ket{H_{i,m,n}}$. For the in-plane coupling, ${L}=\frac{\partial^2}{\partial z^2}$; for the surface coupling, ${L}=-\delta(|d|)\frac{\partial}{\partial z}$.}
\label{atG}
\end{figure}
\begin{figure*}[htbp]
\centerline{
\includegraphics[width=17cm]{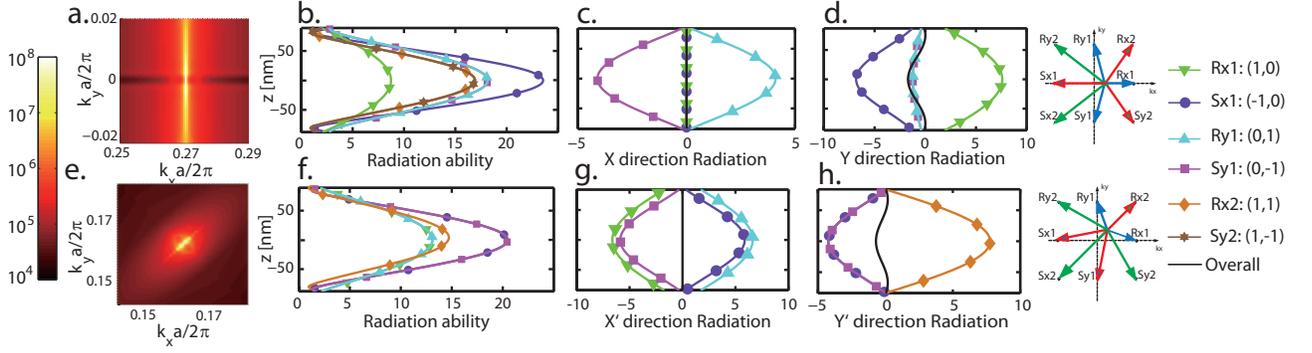}}
 \caption{(Color online). CWT-calculated $Q_{r}$ \textbf{a.} TM$_{1}$ mode near $\Gamma-X$ BIC \textbf{e.} TM$_{2}$ mode near $\Gamma-M$ BIC. CWT-calculated radiation ability at the symmetry (defined by the modulus of the diffraction into the open channel of individual wavevectors with normalized profiles, i.e. $\|\langle G_{i,-m,-n} H_{i,m,n}\rangle\|$, $i=x,y$) for the \textbf{b.} $\Gamma-X$ BIC \textbf{f.} $\Gamma-M$ BIC. Radiation profile decomposition within the slab of \textbf{c.d.} TM$_{1}$ $\Gamma-X$ BIC in the \textbf{x} and \textbf{y} directions \textbf{g.h.} TM$_{2}$ $\Gamma-M$ BIC in the \textbf{x'} (y=x) and \textbf{y'} (y=-x) directions. The radiation profiles are calculated by $P_{m,n}\sum_{m',n'}\bra{\textbf{V}_{TM_{1(2)}}}G_{i,-m',-n'}\ket{H_{i,m',n'}}, i=x,y$, where $\textbf{V}_{TM_{1(2)}}$ is the eigenvector of the $\Gamma-X(M)$ BIC and P$_{mn}$ denotes corresponding projection operators, i.e. $\ket{mn}\bra{mn}$.}
\label{GXDetail}
\end{figure*}
Moreover, the leaky wave (open channel) can be calculated  in terms of the closed channels:
\begin{eqnarray}
H_{i,0,0}=\sum_{\substack{m,n}}\bra{G}\kappa_{\substack{m,n}}\left(\frac{\partial^2}{\partial z^2}-\delta(|d|)\frac{\partial}{\partial z}\right)\ket{H_{i,m,n}},
\label{Open}
\end{eqnarray}
where $G=\left(k_{0}^2+\delta(|d|)\Delta\kappa\frac{\partial}{\partial z}+\kappa_{a}\frac{\partial^2}{\partial z^2}\right)^{-1}$ is the Green function.

The radiative wave includes the contributions from closed channels with different weights and phases (Fig. \ref{schematic}a). We calculate the band structure (with a wave truncation order \cite{LCPRB} of 10 for convergence) near the $\Gamma^2$ point where four band-edge modes are identified as TM$_{1-4}$ in Fig. \ref{atG}a. Moreover, $Q_{r}$ of the TM$_{1}$ mode in the vicinity of $\Gamma^2$ is depicted using the CWT (Fig. \ref{atG}b). Because of the geometric symmetry, all of the coupling coefficients ($\kappa_{mn}$) are symmetric at $\Gamma$ (Fig. \ref{schematic}c), which leads to complete destructive interference.
For TM-like modes, a partial cancellation occurs between the in-plane and surface coupling. As shown in Fig. \ref{atG}c, the strengths of the two coupling mechanisms are comparable in amplitude but opposite in sign, which reduces the overall radiation.

The ``perfect symmetry" picture for at-$\Gamma$ BICs cannot straightforwardly explain the recently reported tunable BICs that occur at seemingly unremarkable $k$ points without symmetry incompatibility. From the analytical perspective of CWT, we found
the participation of other orders of $\Gamma$ points may create new symmetry for the formation of tunable BICs.

For the reported tunable BIC at $k\simeq0.25$ in the $\Gamma-X$ direction \cite{GMRNAT}, the wavevectors of $S_{x1}$, $R_{x2}$, and $S_{y2}$ become degenerate and create a new phase matching $\beta = 1.25\beta_{0}$ via a triangular symmetry (Fig. \ref{schematic}d).
As a result, most of the energy coupled to the radiative open channel can be cancelled, which forms a $\Gamma-X$ BIC (Fig. \ref{GXDetail}a).
Fig. \ref{GXDetail}b shows the radiation ability of the six most significant interfering waves in this case. Among these, $S_{x1}$ has the highest radiation ability while $R_{x2}$ and $S_{y2}$ are comparable. Although $R_{y1}$ and $S_{y1}$ also exhibit considerable contributions; however, the majority of the corresponding radiation cancel with each other because of the symmetry with respect to \textbf{x} axis.

The entire radiation cancellation is demonstrated by mixing all of the participating radiative channels via the mode eigenvectors, as shown in Figs. \ref{GXDetail}c and d. In the \textbf{x} direction, all of the radiative waves are cancelled because of symmetry. In contrast, no apparent symmetry holds in the \textbf{y} direction. As expected, the radiation projected onto $S_{x1}$ is rather large. The other large component $R_{x1}$ comprises the projection of $R_{x2}$, $S_{y2}$, and $R_{x1}$,
in which $R_{x2}$ and $S_{y2}$ dominate because the radiation ability of $R_{x1}$ is rather low (Fig. \ref{GXDetail}b). Therefore, the accidental degenerate wavevectors $S_{x1}$, $R_{x2}$ and $S_{y2}$ contribute dominant radiation, and $S_{x1}$ is in opposite sign with $R_{x2}$ and $S_{y2}$ (Fig. \ref{GXDetail}d). Hence, by incorporating the radiation from all possible channels, the weighted destructive interference causes the overall radiation to be suppressed (close to 0) as shown in Fig. \ref{GXDetail}d.

Interestingly, similar accidental symmetry that may lead to new off-$\Gamma$ BICs also exists elsewhere. For instance, $R_{x2}$, $S_{x1}$, and $S_{y2}$ become degenerate at $k=\sqrt{2}/6$ in the $\Gamma-M$ direction (Fig. \ref{schematic}e), forming another triangular symmetry and a new $\Gamma-M$ BIC (Fig. \ref{GXDetail}e).
Fig. \ref{GXDetail}f shows the radiation ability of the five most significant individual waves: $S_{x1}$ and $S_{y1}$ exhibit the highest radiation ability, and the radiation from $R_{x2}$ is also remarkbly large. As in the $\Gamma-X$ case, the majority of radiation from $R_{x1}$ and $R_{y1}$ is cancelled owing to symmetry (see Fig. \ref{GXDetail}g). Thus, the degenerate wavevectors $S_{x1}$, $S_{y1}$ and $R_{x2}$ contribute dominant radiation.

Choosing the polarizations along the $y=x $ (as $x'$) and $y=-x$ (as $y'$) axes \cite{LCPRB2} allows for better interpretation of the weighted destructive interference of the $\Gamma-M$ BIC after mixing with the TM$_{2}$ eigenvectors (Figs. \ref{GXDetail}g and h). As previously stated, the radiation in the $x'$ direction remains symmetric and cancels out. In the $y'$ direction, the radiation projected onto $S_{x1}$ and $S_{y1}$ are identical and remarkably large, and their sign is opposite to that of the radiation projected onto $R_{x2}$. Hence, the weighted destructive interference can be achieved owing to the accidental triangular symmetry, and the overall radiation is suppressed (close to 0).

The off-$\Gamma$ BICs do not necessarily occur at the exact accidental symmetry point. Unlike the at-$\Gamma$ BICs, the off-$\Gamma$ BICs have different closed-channel weights in their contributions to radiative open channels (determined by $\kappa_{mn}$), and hence, a small amount of residual radiation may remain after cancellation at the exact symmetry point. Nevertheless, the interference can be continuously adjusted using the wavevector such that weighted destructive interference always occurs because of the new symmetry, giving rise to a tunable trapping of light. As long as the mirror-flip and inversion symmetry are kept intact \cite{GMRNAT}, tunable BICs can be stably found.

\begin{figure}[htbp]
\centerline{
\includegraphics[width=8.5cm]{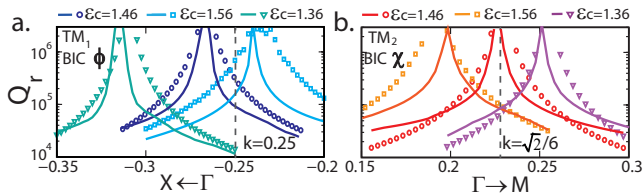}}
 \caption{(Color online). Location shift of tunable BICs in the \textbf{a.} $\Gamma-X$ and \textbf{b.} $\Gamma-M$ direction using CWT (solid lines) and FDTD (dots). $\epsilon_{c}$ is the cladding permittivity. Dashed lines indicate the exact triangular symmetry (Figs. \ref{schematic}d and e).}
\label{QQ}
\end{figure}

Furthermore, the locations of off-$\Gamma$ BICs can be shifted by changing various parameters, such as the cladding permittivity $\varepsilon_{c}$. As shown in Figs. \ref{QQ}a and b, the locations of the off-$\Gamma$ BICs shift away from the $\Gamma$ point with higher index contrast (i.e. smaller cladding permittivity) and vice versa. When the cladding permittivity varies, the out-of-plane profiles of the individual channels change accordingly, modifying the via-the-continuum coupling weights, which leads to the location shifts of the tunable BICs.

\begin{figure}[htbp]
\centerline{
\includegraphics[width=8cm]{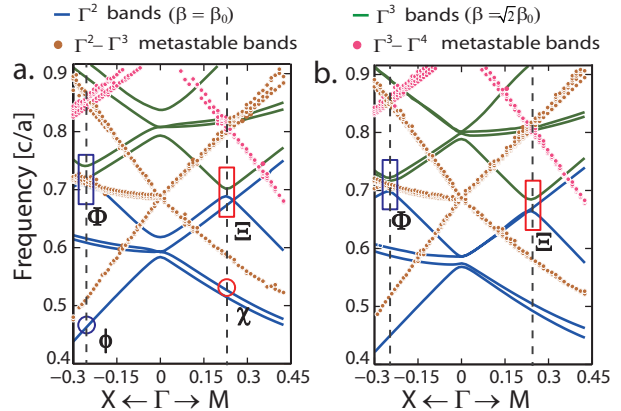}}
 \caption{(Color online). \textbf{a.} TM-like and \textbf{b.} TE-like band structures in the continuum using FDTD. $\phi$ and $\chi$ indicate locations of tunable BICs. $\Phi$ and $\Xi$ indicate bandgaps formed by triangular symmetry. }
\label{FDTD}
\end{figure}

It should be emphasized that the surface coupling plays an important role in the formation of tunable BICs in TM-like modes. The surface coupling in TM-like modes, unlike that in TE-like modes, pre-compensates the majority of the large leakage caused by the in-plane coupling in TM-like modes (Fig. \ref{atG}c). Hence, sufficiently low radiation ability of the uncoupled closed channels is a prerequisite for the formation of tunable BICs.
This criterion also applies for TE-like modes that only possess in-plane coupling, provided that the overall radiation can be effectively suppressed by proper design. For the structure shown in Fig. \ref{schematic}a, TE-like tunable BICs can hardly be found, as the uncoupled radiation of TE-like modes is too large for via-the-continuum channels to compensate. We calculate the continuum region for both TE and TM-like modes as shown in Fig. \ref{FDTD}. Bandgaps $\Phi$ and $\Xi$ appear in both TE and TM-like bands, validating the interference by the accidental new phase matchings (Figs. \ref{schematic}d and e). However, owing to the large radiation ability of the uncoupled channels, TE-like modes can only demonstrate stationary BICs at $k\simeq0$ with symmetry incompatibility. The tunable BICs $\phi$ (Fig. \ref{QQ}a) and $\chi$ (Fig. \ref{QQ}b) solely occur in TM-like modes.

Thus, the formation of tunable BICs is quite clear: the partial cancellation of the in-plane and the surface coupling ensures rather low radiation ability of the separate closed channels. New accidental symmetry induces strong coupling in via-the-continuum channels (but still far weaker than the pre-cancellation within single closed channels). Finally, scanning of the wavevector allows achieving the weighted destructive interference between the closed and open channels. The via-the-continuum coupling also exists for 1D situations \cite{yamamotoTE}. According to the similar phase matching criteria for 2D cases, the 1D case should potentially possess an off-$\Gamma$ BIC in the vicinity of 0.5\textbf{G}, which was reported in \cite{Pho2}. Therefore, both fixed BICs \cite{Pho1,GMRPRL} and tunable BICs \cite{Pho2,GMRNAT,LSA} are realizable in 1D and 2D structures.

The off-$\Gamma$ BICs can also be understood as the interband coupling \cite{InterPRB} between the $\Gamma^2$ and $\Gamma^3$ bands \cite{VecBeam}. With infinitely thick PhC, the coupling between different $\Gamma$ orders is forbidden because of orthogonality. However, for PhC with finite thickness, the orthogonality is broken, and interband coupling is allowed owing to the indirect via-the-continuum coupling within near fields. The destructive interference of the via-the-continuum coupling induces BICs, while the constructive interference forms metastable bands (Fig. \ref{FDTD}). Because the mixed metastable modes are not supported by the PhC periodicity, they quickly dissipate or couple back to the stable bands, rendering their low $Q_{r}$ in the entire Brillouin zone.

The discussion of BICs in this letter focuses on the cancellation of the out-of-plane radiation but neglecting the in-plane loss. For a PhC slab with a finite area, energy may leak at planar boundaries, causing more loss. The in-plane loss may be negligible for a macroscopic PhC slab \cite{GMRPRL}, but it must be considered in realizing a three-dimensional BIC within a smaller area. Thus, the planar permittivity (i.e., potential) distribution should be addressed to reduce the considerable in-plane loss. For photonics, we believe a damped oscillating planar permittivity envelope is a promising solution, which corresponds to the initial proposal for a quantum system \cite{von}.

In this Letter, we provided an analytical perspective for BICs in PhC slabs. For separate closed channels, the compensation of in-plane and surface coupling makes TM-like modes easier to trap. By tuning the wavevector, the via-the-continuum coupling can form weighted destructive interference through new accidental symmetry.

\begin{acknowledgments}
This work was supported by the National Key Basic Research Program of China (973 Program of China) 2013CB329205 and the National Natural Science Foundation of China (NSFC) under grant No.61307089.
\end{acknowledgments}

\bibliographystyle{apsrev4-1}
\bibliography{Trapped_PRL}

\end{document}